\documentclass[journal,article,submit,moreauthors,dvipdfm,12pt,a4paper]{mdpi} 
\lastpage{x}
\doinum{10.3390/------}
\pubvolume{xx}
\pubyear{2012}
\history{Received: xx / Accepted: xx / Published: xx}

\Title{A Phase Space Diagram for Gravity}

\Author{X. Hernandez $^{1,\star}$}

\address[1]{%
$^{1}$ Instituto de Astronom\'{\i}a, Universidad Nacional Aut\'onoma de M\'exico, AP 70-264, Distrito Federal 04510, M\'exico}

\corres{xavier@astro.unam.mx.}

\abstract{ 
In modified theories of gravity including a critical acceleration scale, $a_{0}$, 
a critical length scale, $r_{M}=(GM/a_{0})^{1/2}$, will naturally arise, 
with the transition from the Newtonian to the dark matter mimicking regime occurring 
for systems larger than $r_{M}$. This adds a second critical scale to gravity, in 
addition to the one introduced by the criterion $v < c$ of the Shwarzschild radius, 
$r_{S}=2GM/c^{2}$. The distinct dependencies of the two above length scales give 
rise to non-trivial phenomenology in the (mass, length) plane for astrophysical 
structures, which we explore here. Surprisingly, extrapolation to atomic scales
suggests gravity should be at the dark matter mimicking regime there.
}


\begin{document}

\section{Introduction}
 
Over the past years numerous approaches have appeared, proposing to interpret 
galactic rotation curves and other dynamical astrophysical observations, as well as
gravitational lensing, as
gravitational effects. All the many theoretical approaches proposed as alternatives
to dark matter share in common, by construction, the reproduction in the relevant 
acceleration and velocity limits, of accelerations which tend to $\propto M^{1/2}/r$ 
for large distances from a spherical mass $M$. This regime appears at accelerations
lower than the critical value of the MOND $a_{0}$ parameter, to reproduce the observed
flat rotation curves and Tully-Fisher relation of spiral galaxies. Examples of the
above are the modified dynamics approach of MOND e.g. \cite{Mil83}, the Lagrangian 
MOND schemes of e.g. \cite{Zha10}, covariant MOND formulations
e.g. \cite{Wu08}, \cite{Ber11}, the Tensor-Vector-Scalar formalism of TeVeS of \cite{Bek04}, or
conformal gravity theories e.g. \cite{Man89}.

From a cosmological perspective, the qualitative similarity between the early inflationary
phase and the current late accelerated expansion phase, has been interpreted as evidence
for a common physical origin for both, in terms of modified gravity, \cite{Noji03}.
This approach has been extensively explored over the past years by several authors, 
who have now showed the consistency of the proposal with all global expansion history
observations, for a variety of extensions to general relativity e.g. \cite{Noji05}, \cite{Noji07}, 
\cite{Eli04}, \cite{Eli05}, \cite{Cog05}, \cite{Cog06}, \cite{Cog07}. The connection between 
such approaches and dark matter inferences at galactic dynamics level has also been explored 
for the case of $F(R)$ modifications to general relativity by e.g. \cite{Cap07}, \cite{Cap11}, 
\cite{Nap12}.

Very recently, independent observations for three distinct types of astrophysical systems
have severely challenged the standard gravity plus dark matter scenario, showing
a phenomenology which is actually what modified gravity theories predict. Firstly,
the case of globular clusters is interesting, with these systems having traditionally 
been thought of as classical examples of purely Newtonian gravity, and exhibiting values 
of $a>a_{0}$ throughout most of their extent. Recently however, results sampling stellar 
kinematics in their outskirts by \cite{Sca10} and \cite{Sca11} have indicated the appearance 
of MOND type dynamics appearing precisely beyond the point where accelerations fall below $a_{0}$. 
Further, the recent analysis of \cite{HJi12} has shown the appearance of a ``Tully-Fisher'' 
relation in these systems, a scaling of their dispersion velocities at the outskirts 
with the fourth root of their total masses. These observations are precisely what is 
expected under modified theories of gravity, but would require rather contrived and fine 
tuned explanations under standard gravity.
Secondly, \cite{HJiA12} recently reported a gravitational anomaly of the type generally
ascribed to dark matter at galactic scales, at the much smaller and unexpected scales
of 1-10 pc associated with wide binaries in the solar neighbourhood. These authors show
that typical binary stellar orbital velocities cease to fall with separation along Keplerian
expectations, and settle at a constant value consistent with modified gravity predictions, 
exactly on crossing separations where the acceleration falls below $a_{0}$.
Finally, \cite{Lee10} showed that the inferred infall velocity of the bullet cluster
is inconsistent with the standard cosmological scenario, where much smaller limit 
encounter velocities are expected at those redshifts. The problem was more carefully 
re-analysed by \cite{Tho12}, reaching identical conclusions. The inconsistency 
stems from the physically imposed escape velocity limit present in standard gravity;
the ``bullet'' should not hit the ``target'' at more than the escape velocity of the
joint system, as it very clearly did. The slower radial fall-off of the gravitational 
force in modified gravity schemes however, makes it natural to obtain encounter 
velocities much beyond classical escape velocities, as shown in e.g. \cite{Mof10}.
The above mentioned observations put us in a situation where modifications to gravity
at low acceleration scales cease to be a matter of choice, and now appear inevitable.

The introduction of a critical acceleration in modified gravity theories in turn implies the appearance of a 
critical mass dependent length scale $r_{M}=(GM/a_{0})^{1/2}$. Systems having extents 
larger than their corresponding $r_{M}$ values will be in the dark matter mimicking 
regime, while those smaller than this value will be in the Newtonian regime, provided 
equilibrium velocities satisfy $v<<c$, e.g. \cite{Mil08}, \cite{Her10}. The appearance of 
a critical mass dependent length scale is not new to gravity, in the covariant version 
of Newtonian gravity, general relativity, the introduction of a critical velocity, $c$, 
introduces a corresponding critical mass dependent length scale $r_{S}=2GM/c^{2}$.

Generally, we are accustomed to thinking in terms of Shwarzschild radii for astrophysical
objects which are much smaller than the extent of the systems in question, which can then be
smaller or larger than their corresponding $r_{M}$ values, depending on whether observed
dynamics satisfy Newtonian expectations or not. A general consistency check for the 
gravitational interpretation of astrophysical dynamics is found in that not a single high 
acceleration system ($a>a_{0}$) is known where dark matter is required, and conversely,
not a single low acceleration system ($a<a_{0}$) is known where dark matter is not
required, when interpreting observations under Newtonian gravity. An exception
to either of the two above rules would seriously challenge many of the modified
theories of gravity currently under consideration.

Going back to the usual hierarchy $r_{S}<<r<r_{M}$ or $r_{S}<<r_{M}<r$ for astrophysical
objects in the Newtonian or dark matter mimicking regimes, we note that the distinct
mass scalings of $r_{S}$ and $r_{M}$ imply that at sufficiently large masses the situation
$r_{M}<r_{S}$ could arise. This leads to non-trivial structure in the (mass, radius)
plane for astrophysical objects, which we explore across 25 orders in magnitude in both axis
in the following section. Section 3 then presents the results of extrapolating 
the empirical phenomenology of astrophysical scales down to the atomic regime, 
with the interesting prediction that at those scales, gravity should appear to be at 
the dark matter mimicking regime. Finally, our conclusions are presented in section 4.


\section{A Gravitational Phase Space Diagram}

We begin by examining the distinct dependencies of the two critical length
scales which will appear in any covariant theory of gravity aiming at explaining 
the observed astrophysical phenomenology at galactic scales, without invoking dark matter:

\begin{equation}
r_{S}=\frac{2 G M}{c^{2}},
\end{equation}

and

\begin{equation}
r_{M}=\left( \frac{G M}{a_{0}} \right)^{1/2}.
\end{equation}

It is now obvious that a critical dimensionless parameter of the problem will be
the ratio of the above two radii, $b=r_{S}/r_{M}$. This parameter will 
be very small for most astrophysical objects.
Whilst $r_{S}$ scales with $M$, $r_{M}$ scales only with $M^{1/2}$. This
implies a reversal of the accustomed hierarchy $r_{S}<<r_{M}$ into $r_{M}<r_{S}$ at
sufficiently large masses, when $b$ will transit from $b<1$ to $b>1$, with a critical
point appearing at $b=1$.

To better appreciate the distinct regions which will appear in the 
(mass, radius) plane, we plot figure 1, where the two thick solid lines show the two physical
critical conditions $v<c$ and $a=a_{0}$, and their corresponding resulting mass dependent
length scales of equations (1) and (2), in a log-log scale. The dashed line below the
$r=r_{S}$ condition gives the region where relativistic effects begin to appear, at the 
threshold where equilibrium velocities cease to be negligible with respect to $c$, 
of order $v=0.01 c$.

We see that the Newtonian region is restricted to a wedge extending downwards and limited 
along the top by the dotted line $r=100r_{S}$, and from the lower side by the condition
$r=r_{M}$. This clearly encompasses gravity at the planetary scale, the solar system, 
globular clusters (excluding their outer regions), while binary stars transit from 
this region to the modified gravity regime to the right of it.
Elliptical galaxies appear somewhat at the edge of this region. Indeed, 
in \cite{Men11} some of us showed how the observed scaling relations for ellipticals, 
along with most of the tilt in the fundamental plane, can be easily explained by the appearance 
of non Newtonian effects outside their core regions, in consistency with the approach of the 
$a=a_{0}$ threshold.

\begin{figure}
\includegraphics[width=16.0cm,height=12.0cm]{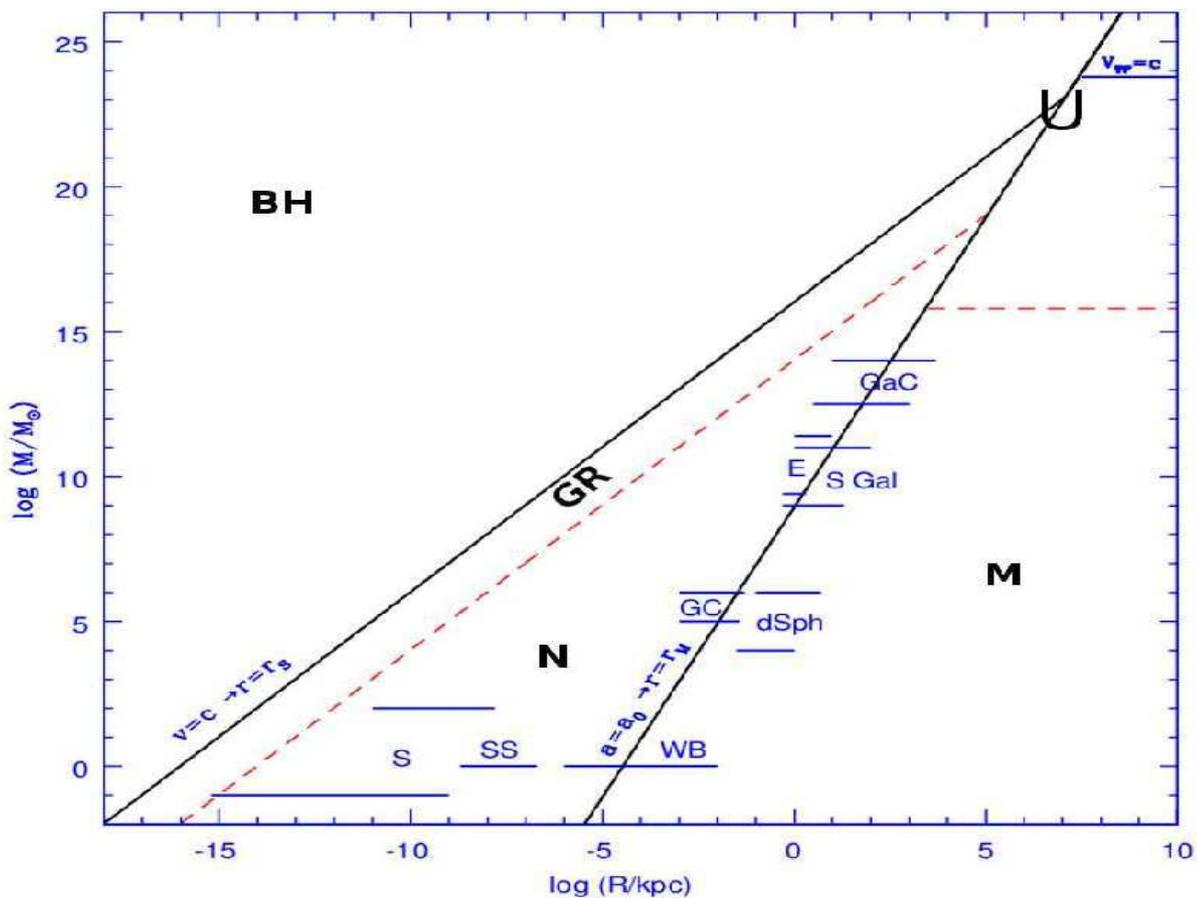}
\caption{Phase space diagram for self gravitating equilibrium configurations. The labelled solid
lines give the mass dependant scale radii resulting from the two limit conditions $v=c$ and
$a=a_{0}$, $r_{S}=2GM/c^{2}$ and $r_{M}=(GM/a_{0})^{1/2}$. The approach to the former
from below, signals the relativistic region, whilst the approach to latter from the left, 
denotes the transition from the Newtonian to the dark matter mimicking regime. The labels
identify the regions occupied by different astrophysical objects; the solar system, SS, 
stars, S, wide binaries, WB, globular clusters, GC, dwarf spheroidal galaxies, dSph, elliptical galaxies, E, spiral 
galaxies, S Gal and galaxy clusters, GaC. Distinct regions of the diagram are labelled; black holes, BH,
appearance of relativistic effects, GR, the Newtonian region, N, the modified gravity regime, M, and the
critical density of the universe, or the dark energy density, coinciding with the critical point b=1 where $r_{S}=r_{M}$.}
\end{figure}

Within this Newtonian wedge region, increasing the mass or reducing the radius drives a system into the 
relativistic region, and then into the black hole regime. Conversely, reducing the mass
or increasing the radius shifts an object from the Newtonian regime into the dark matter
mimicking region, for example, in going from globular clusters to dwarf spheroidal galaxies (dSphs),
objects with comparable masses, but qualitatively distinct dynamics. The details of the transition
are open to debate, and are commonly expressed in terms of the choice of the corresponding MOND $\mu$ 
transition function, e.g. \cite{Mil83}, \cite{Fam05}, \cite{Zha07}, \cite{Men11}, \cite{HJi12}, \cite{Qas11}.

To the right of the $r=r_{M}$ line we see the dark matter mimicking region, occupied for example
by the dSph galaxies, the most "dark matter dominated" systems known. These have mass to light ratios
sometimes in the thousands, under Newtonian interpretations, e.g \cite{Gil07}. Spiral
galaxies transit from being to the left of this line in their internal regions, to being to 
the right of it as one moves away along their disks. For the Milky Way, we see the Solar
radius appearing slightly to the right of the $r=r_{M}$ threshold, consistent with the Newtonian 
deduction of a 50\% dark matter content within this radius, e.g. \cite{Men11}. We see also galaxy clusters to
the right of the line marking the end of the Newtonian region at their outskirts.

At a very large critical mass of $ M_{b} = 5.06 \times 10^{23} M_{\odot}$, we see the intersection of the two
gravitational critical radii at $b=1$. The corresponding radius being of $R_{b} = 2.5 \times 10^{4} Mpc$. 
The above critical mass and radius are essentially the only such quantities which can be constructed dimensionally
from $G$, $c$ and $a_{0}$, $M_{c}=c^{4}/(G a_{0})$ and $R_{c}=c^{2}/a_{0}$. It
is interesting to note that the density which corresponds to $b=1$ critical parameters,
$\rho_{b}=M_{b}/R_{b}^{3} = 4.8 \times 10^{-27} kg m ^{-3}$ agrees to within a factor of 2 with the 
critical density of the universe of $\rho_{c}=8.4 \times 10^{-27} kg m^{-3}$ or equivalently, the
density of ``dark energy'' inferred under GR. This points to the appearance of the relativistic 
dark matter mimicking region at cosmological densities somewhat lower than those present today, 
coinciding with the regime where the accelerated expansion of the Universe is detected.
Thus, we see that the critical density of the universe is also critical in the sense of $b=1$.
Since $H_{0}^{2}=8 \pi G \rho_{c} /3$, the agreement of $\rho_{b} \approx \rho_{c}$ is equivalent
to the well known numerical coincidence of $a_{0} \approx H_{0} c$ (e.g. \cite{BeCa11}),  
and indeed, could point to the physical origin of the numerical equivalence in question.

Also, the end of the Newtonian sector at masses of order $10^{18} M_{\odot}$ implies a region
where the dark matter mimicking regime transits directly into the relativistic one, without passing
through a Newtonian region. To the right of the $r=r_{M}$ threshold, equilibrium velocities
satisfy the Tully-Fisher relation:

\begin{equation}
\left( \frac{V}{c} \right)^{2}=\left( \frac{G M a_{0}}{c^{4}}\right)^{1/2}=\left( \frac{M}{M_{c}} \right)^{1/2}.
\end{equation}

However, this scaling can not continue to be the case for arbitrarily large masses, which would 
imply equilibrium velocities larger than $c$. We must therefore think of a modification of the type


\begin{equation}
\left( \frac{V}{c} \right)^{2} =\left( \frac{M}{M + M_{c}} \right)^{1/2}.  
\end{equation}

The thin horizontal continuous line gives the limit mass $M=M_{c}$, 
the level at which the dark matter mimicking regime becomes relativistic. The corresponding threshold
at which this regime should begin to exhibit relativistic effects, where the standard Tully-Fisher
relation yields $V=0.01c$, is given by the horizontal dashed line. We see that galaxy clusters
lie very close to this line, in fact, dispersion velocities in clusters of galaxies
often exceed $1000 km/s$, much more than the values of around $50 km/s$ of the orbit of Mercury,
where relativistic effects begin to become apparent. This alerts to the fact that galaxy clusters
probably present non-negligible relativistic effects, and can not be treated under non-relativistic 
modified gravity schemes. This appears obvious from the region occupied by galaxy clusters in 
figure (1), only slightly below the horizontal dotted line mentioned.

In view of the above, it is probably more correct to think of the relativistic regime, which 
within the Newtonian region is defined by the dashed line $r=100r_{S}$, as blending 
continuously into the horizontal dashed line appearing a little below $M=10^{16} M_{\odot}$. Above the
corresponding $v=c$ line one can speculate about "MONDian" black holes and other 
phenomena, but in the absence of observations, we restrict the discussion to the regions probed
by known astrophysical objects. The relativistic "MONDian" regime, $v\sim c$ and $a<a_{0}$
appears populated only by the critical density of the universe.


\section{Extrapolation to Atomic Scales}

It is interesting to calculate on what side of the $r=r_{M}$ divide systems at the
atomic scale lie. Taking $M=N m_{p}$ with $m_{p}$ the proton mass, it is immediate
to calculate from equation(2) $r_{M}=3.05 \times 10^{-4} N^{1/2} \AA$, with a resulting 
value of $b=7.44 \times10^{-32} N^{1/2}$. Therefore atoms, systems in the $\AA$ range of scales
with $N$ of order a few,
lie several orders of magnitude to the right of the $r=r_{M}$ divide, as is the case of 
galactic systems. Hence, the extrapolation of gravitational phenomenology under modified gravity ideas, 
implies gravity at atomic scales will be at the dark matter mimicking regime. 

One should therefore expect that at atomic scales, a test mass in the presence of a 
much larger mass $M$, will experience a gravitational attraction several thousand 
times larger than the Newtonian prediction. This remains many orders of magnitude
below the electromagnetic effects, which obviously still largely dominate.
However, if such effects can be accounted for, a residual force per unit mass should 
appear given by:

\begin{equation}
F=\frac{c^{2} b}{2 r},
\end{equation}

\noindent with a corresponding potential $\Phi=(c^{2} b/2) ln(r/r_{S})$, where $r_{S}$ has been introduced 
for dimensional consistency. We see again the critical parameter $b$ appearing. 
This force will be several orders of magnitude larger than the
Newtonian value. It is important to notice that this prediction is generic to many modified
gravity theories, which explain the dynamics otherwise ascribed to dark matter as gravitational
effects, largely independent of the details of the covariant framework behind the observed phenomenology.
The above expectations could be relevant i light of forthcoming micro-gravity experiments, e.g.
the forthcoming ESA STE-QUEST satellite.


\section{Conclusions}

We have shown that since in modified theories of gravity reproducing the observed astrophysical
phenomenology a second gravitational mass dependent length scale appears in addition
to the  Shwarzschild radius, non-trivial structure appears in a (mass-radius) phase space
diagram for gravity.

The disappearance of the Newtonian region for masses slightly above galactic cluster scales
identifies a limit above which low velocity dark matter mimicking phenomenology can transit
into its relativistic regime, without an intermediary Newtonian region.

The coincidence of the critical mass and radius at this point with the critical density
of the universe could be interpreted as a clue towards understanding the recent appearance
of the accelerated expansion of the Universe, within the framework of modified theories
of gravity in general.

In going to the smallest scales available to direct experimentation, we see that a 
prediction appears, in the form of gravity at atomic level being decidedly at the dark
matter mimicking regime. This constitutes an exciting prediction for future micro-gravity
experiments.


\section*{Acknowledgements}
 Xavier Hernandez acknowledges financial support from UNAM-DGAPA grant IN103011.

\bibliographystyle{mdpi}
\makeatletter
\renewcommand\@biblabel[1]{#1. }
\makeatother

\begin{thebibliography}{99}



\bibitem{Mil83} 
Milgrom, M. A modification of the Newtonian dynamics as a possible alternative to the hidden mass hypothesis. 
{\em Astrophys. J.}, {\bf 1983}, {\em 270}, 365-370.

\bibitem{Zha10}
Zhao, H.; Famaey, B. Comparing different realizations of modified Newtonian dynamics: Virial theorem and elliptical shells.	
{\em Phys. Rev. D}, {\bf 2010}, {\em 83}, 087304-087308.

\bibitem{Wu08} 
Wu, X.; Famaey, B.; Gentile, G.; Perets, H.; Zhao, H. S. Milky Way potentials in cold dark matter and MOdified Newtonian Dynamics. Is the Large Magellanic Cloud on a bound orbit?.
{\em Mon. Not. R. Astron. Soc.}, {\bf 2008}, {\em 386}, 2199-2208.

\bibitem{Ber11}
Bernal, T.; Capozziello, S.; Hidalgo, J. C.; Mendoza, S. Recovering MOND from extended metric theories of gravity.
{\em  Eur. Phys. J. C},{\bf 2011},{\em 71}, 1794-1801. 

\bibitem{Bek04} 
Bekenstein, J. D. Relativistic gravitation theory for the modified Newtonian dynamics paradigm.
{\em Phys. Rev. D}, {\bf 2004}, {\em 70}, 083509-083537.

\bibitem{Man89} 
Mannheim, P. D.; Kazanas, D. Exact vacuum solution to conformal Weyl gravity and galactic rotation curves.
{\em Astrophys. J.}, {\bf 1989}, {\em 325} 635-638.




\bibitem{Noji03}
Nojiri, S.; Odintsov, S. D.; Modified gravity with negative and positive powers of curvature: Unification of inflation and cosmic acceleration.
{\em Phys. Rev. D}, {\bf 2003}, {\em 68}, 123512-123522.


\bibitem{Noji05}
Nojiri, S.; Odintsov, S. D.; Tsujikawa, S. Properties of singularities in (phantom) dark energy universe.
{\em Phys. Rev. D}, {\bf 2005}, {\em 71}, 063004-063020. 



\bibitem{Noji07}
Nojiri, S.; Odintsov, S. D. Introduction to modified gravity and gravitational alternative for dark energy.
{\em Int. J. Geom. Meth. Mod. Phys.}, {\bf 2007}, {\em 4}, 115-146.


\bibitem{Eli04}
Elizalde, E.; Nojiri, S.; Odintsov, S. D. Late-time cosmology in (phantom) scalar-tensor theory: Dark energy and the cosmic speed-up.
{\em Phys. Rev. D}, {\bf 2004}, {\em 70}, 043539-043559.


\bibitem{Eli05}
Elizalde, E.; Nojiri, S.; Odintsov, S. D.; Wang, P. Dark energy: Vacuum fluctuations, the effective phantom phase, and holography.
{\em Phys. Rev. D}, {\bf 2005}, {\em 71}, 103504-103512.



\bibitem{Cog05}
Cognola, G.; Elizalde, E.; Nojiri, S.; Odintsov, S. D.; Zerbini, S. One-loop f(R) gravity in de Sitter universe.
{\em JCAP}, {\bf 2005}, {\em 02}, 010-037.


\bibitem{Cog06}
Cognola, G.; Elizalde, E.; Nojiri, S.; Odintsov, S. D.; Zerbini, S. Dark energy in modified Gauss-Bonnet gravity: Late-time acceleration and the hierarchy problem.
{\em Phys. Rev. D}, {\bf 2006}, {\em 73}, 084007-084023.


\bibitem{Cog07}
Cognola, G.; Elizalde, E.; Nojiri, S.; Odintsov, S. D.; Sebastiani, L.; Zerbini, S. A Class of viable modified f(R) gravities describing inflation and the onset of accelerated expansion.
{\em Phys. Rev. D}, {\bf 2008}, {\em 77}, 046009-046020.




\bibitem{Cap07} 
Capozziello, S.; Cardone, V. F.; Troisi, A. Low surface brightness galaxy rotation curves in the low energy limit of R$^n$ gravity: no need for dark matter?
{\em Mon. Not. R. Astron. Soc.}, {\bf 2007}, {\em 375}, 1423-1440.

\bibitem{Cap11} 
Capozziello, S.; De Laurentis, M. Extended Theories of Gravity.
{\em Phys. Rep.}, {\bf 2011}, {\em 509}, 167-321.

\bibitem{Nap12} 
Napolitano, N. R.; Capozziello, S.; Romanowsky, A. J.; Capaccioli, M.; Tortora, C. Testing Yukawa-like potentials from f(R)-gravity in elliptical galaxies.
{\em Astrophys. J.}, {\bf 2012}, {\em in press}, arXiv:1201.3363.







\bibitem{Sca10} 
Scarpa, R.; Falomo, R. Testing Newtonian gravity in the low acceleration regime with globular clusters: the case of ω Centauri revisited.
{\em Astron. Astrophys.}, {\bf 2010}, {\em 523}, A43-A49.

\bibitem{Sca11} 
Scarpa, R.; Marconi, G.; Carraro, G.; Falomo, R.; Villanova, S. Testing Newtonian gravity with distant globular clusters: NGC 1851 and NGC 1904.
{\em Astron. Astrophys.}, {\bf 2011}, {\em 525}, A148-A158.

\bibitem{HJi12}
Hernandez, X.; Jimenez, M. A. The outskirts of globular clusters as modified gravity probes.
{\em Astrophys. J.}, {\bf 2012}, {\em in press}, arXiv:1108.4021.

\bibitem{HJiA12}
Hernandez, X.; Jimenez, M. A.; Allen, C. Wide binaries as a critical test of classical gravity.
{\em Eur. Phys. J. C}, {\bf 2012}, {\em 72}, 1884-1890.

\bibitem{Lee10}
Lee, J.; Komatsu, E. Bullet Cluster: A Challenge to ΛCDM Cosmology.
{\em Astrophys. J.}, {\bf 2010}, {\em 718}, 60-65.

\bibitem{Tho12}
Thompson, R.; Nagamine, K. Pairwise velocities of dark matter haloes: a test for the Λ cold dark matter model using the bullet cluster.
{\em Mon. Not. R. Astron. Soc.}, {\bf 2012}, {419}, 3560-3570.

\bibitem{Mof10}
Moffat, J. W.; Toth, V. T. Can Modified Gravity (MOG) explain the speeding Bullet (Cluster)?
{\bf 2010},  arXiv:1005.2685.

\bibitem{Mil08} 
Milgrom, M.; Sanders, R. H. Rings and Shells of ``Dark Matter'' as MOND Artifacts.
{\em Astrophys. J.}, {\bf 2008}, {\em 678}, 131-143.

\bibitem{Her10} 
Hernandez, X.; Mendoza, S.; Suarez, T.; Bernal, T. Understanding local dwarf spheroidals and their scaling relations under MOdified Newtonian Dynamics.
{\em Mon. Not. R. Astron. Soc.}, {\bf 2010}, {\em 514}, A101-A109.

\bibitem{Men11} 
Mendoza, S.; Hernandez, X.; Hidalgo, J. C.; Bernal, T. A natural approach to extended Newtonian gravity: tests and predictions across astrophysical scales.
{\em Mon. Not. R. Astron. Soc.}, {\bf 2011}, {\em 411}, 226-234.

\bibitem{Fam05} 
Famaey, B.; Binney, J. Modified Newtonian dynamics in the Milky Way.
{\em Mon. Not. R. Astron. Soc.}, {\bf 2005}, {\em 363}, 603-608.

\bibitem{Zha07} 
Zhao, H. S. Coincidences of Dark Energy with Dark Matter: Clues for a Simple Alternative?
{\em Astrophys. J.}, {\bf 2007}, {\em 671}, L1-L4.

\bibitem{Qas11} 
Qasem, E. Lunar system constraints on the modified theories of gravity.
{\bf 2011}, arXiv:1112.4652.

\bibitem{Gil07} 
Gilmore, G.; Wilkinson, M. I.; Wyse, R. F. G.; Kleyna, J. T.; Koch, A.; Evans, N. W.; Grebel, E. K.
The Observed Properties of Dark Matter on Small Spatial Scales.
{\em Astrophys. J.}, {\bf 2007}, {\em 663}, 948-959.

\bibitem{BeCa11}
Bernal, T.; Capozziello, S.; Cristofano, G.; de Laurentis, M. Mond's Acceleration Scale as a Fundamental Quantity.
{\em Mod. Phys. Lett. A}, {\bf 2011}, {\em 26}, 2677-2687. 








\end{thebibliography}

\end{document}